\documentclass[12pt,english,twoside]{article}

\usepackage[T1]{fontenc}
\usepackage[utf8]{inputenc}
\usepackage[letterpaper, margin=2cm]{geometry}
\usepackage{color}
\usepackage{graphicx}
\usepackage{subfigure}
\graphicspath{{graphics_fold/}}
\usepackage{epstopdf} 
\usepackage{times}
\usepackage{etoolbox}
\usepackage{amsmath} 
\usepackage{amssymb}  
\usepackage{fancyhdr}
\usepackage{psfrag}
\usepackage{enumitem}
\usepackage{indentfirst}       
\usepackage{titlesec}          
\usepackage{hyperref}
\usepackage{babel}
\usepackage{listings}
\usepackage{booktabs}

\addto\captionsenglish{}

\usepackage[colorinlistoftodos]{todonotes} 

\setlist{nolistsep}

\makeatletter
\renewcommand*{\@biblabel}[1]{\hfill#1.}
\makeatother

\makeatletter
\patchcmd{\headrule}{\hrule}{\color{white}\hrule}{}{}
\patchcmd{\footrule}{\hrule}{\color{white}\hrule}{}{}
\makeatother

\fancypagestyle{firstpage}{
\fancyhead{}
\fancyfoot{}
\fancyfoot[LE,RO]{\small \thepage}

}

\makeatletter
\def\maketitle{
  \thispagestyle{firstpage}
  {
   \fontsize{11}{20}\selectfont\rmfamily{} \noindent 
   {}\\

   \smallskip
   \fontsize{17}{20}\selectfont\sffamily{}  \noindent \MakeUppercase{\textbf{\@title}}

   \bigskip
   \fontsize{14}{20}\selectfont\rmfamily{} \noindent \@author
  }
}
\makeatother

\titleformat{\section}
  {\fontsize{14}{14}\normalfont\sffamily\bfseries}
  {\thesection.}{1em}{}                

\titleformat{\subsection}
  {\normalfont\sffamily\bfseries}
  {\thesubsection}{1em}{}

\titleformat{\subsubsection}
  {\normalfont\sffamily\small}
  {\textit{\thesubsubsection}}{1em}\textit{{}}

\usepackage{tikz}
\usepackage{mathrsfs}

\usetikzlibrary{positioning}

\tikzset{%
  every neuron/.style={
    circle,
    draw,
    minimum size=1cm
  },
  neuron missing/.style={
    draw=none, 
    scale=2,
    text height=0.3cm,
    execute at begin node=\color{black}$\vdots$
  },
}

\title{PARAMETRIC OPTIMIZATION OF VIOLIN TOP PLATES USING MACHINE LEARNING}

\author{Davide Salvi, Sebastian Gonzalez, Fabio Antonacci and Augusto Sarti\\
{\small \textit{Dipartimento di Elettronica Informazione e Bioingegneria, Politecnico di Milano, Milan, Italy\\
e-mail: davide.salvi@polimi.it}}}

\begin{document}

\maketitle
\renewcommand{\abstractname}{\vspace{-\baselineskip}} 

\begin{abstract}	\noindent
We recently developed a neural network that receives as input the geometrical and mechanical parameters that define a violin top plate and gives as output its first ten eigenfrequencies computed in free boundary conditions. 
In this manuscript, we use the network to optimize several error functions, with the goal of analyzing the relationship between the eigenspectrum problem for violin top plates and their geometry.
First, we focus on the violin outline. Given a vibratory feature, we find which is the best geometry of the plate to obtain it. Second, we investigate whether, from the vibrational point of view, a change in the outline shape can be compensated by one in the thickness distribution and vice versa. Finally, we analyze how to modify the violin shape to keep its response constant as its material properties vary.
This is an original technique in musical acoustics, where artificial intelligence is not  widely used yet. It allows us to both compute the vibrational behavior of an instrument from its geometry and optimize its shape for a given response.
Furthermore, this method can be of great help to violin makers, who can thus easily understand the effects of the geometry changes in the violins they build, shedding light on one of the most relevant and, at the same time, less understood aspects of the construction process of musical instruments.\\

\end{abstract}

\quad\rule{425pt}{0.4pt}

\section{Introduction}

The question of which is the best geometry for a violin is still open in the instrument making field. We do not have a real clue why Guarneris have one shape and Stradivaris has another, which is the best outline to obtain a given vibrational feature or how a shape change affects the final response of the instrument.
Some studies have been done in this direction \cite{gough2015violin}, and we now know the influence of the shape of the f-holes on the emitted sound \cite{nia2015evolution} or how to change the violin arching to compensate variations of the wood parameters \cite{tinnsten2002numerical}. However, it is not yet clear how the sound is affected by geometric changes of the instrument.

As a first step in understanding this, we focus on the vibrational response of the free violin top plates. This comes from the assumption that the sound radiated by the complete instrument depends, probably in a non-linear and very complex manner, on how the individual elements of the violin vibrate and that the top is the most resonant part of the instrument. 
We developed a method to parametrize a violin top plate in its geometrical and mechanical aspects. With the method, we created a large set of plates that we used to train a neural network to predict the vibratory response of the plate. In \cite{gonzalez2021data} we mainly focused on its predictive power and the inherent correlations between geometry and vibrational response. In this manuscript we focus on several use cases of the network for violin design and optimization.  

In the state of the art, the most used simulation method to study real musical instruments is the Finite Element Method (FEM) \cite{torres2020exploring}, which is a very powerful but also time-consuming process.
The main advantage of our approach lies in the fact that, using neural networks, we can compute and optimize the vibrational response of a violin top plate in few seconds, taking much less time than with a FEM analysis. It also shows how advances in machine learning can serve the development of a 300 years old craft and is the first step to build a network that predicts how a violin sounds from its material parameters and geometry, which could be an invaluable tool for violin makers.

\section{Definition of the datasets and the neural network}

The first step of our study is building the dataset used to train and test the neural network. We do so by creating a wide range of parametrically varied violin top plates and computing their eigenfrequency values in free boundary conditions.
We build the meshes and compute their vibrational responses as in \cite{gonzalez2020a}.
As a starting point for constructing the dataset, we consider a real historical violin that we had the possibility to scan. The reference geometric parameters are those that best fit the top plate of this instrument.
We vary the characteristics of the built plates from three different parameter spaces: shape of the outline, thickness distribution, and mechanical properties.
Each of these is controlled by a different number of variables, which are 20 for the outline ($p_i$), 8 for the thickness ($t_i$), and 7 for the material ($m_i$).
The variation of each parameter is computed in a random way using a zero-mean Gaussian distribution. The full explanation of the dataset creation can be found in \cite{gonzalez2021data}.
Depending on the number of aspects we decide to vary, we build several datasets, where a different number of parameters is needed to define every single mesh.

For what concerns the neural network, we use a feed-forward network \cite{carrasquilla2017machine} with a single hidden layer and a sigmoid activation function connected to a linear output layer.
The number of neurons in the input and hidden layers changes according to the dataset we use, while the architecture of the network remains unchanged.
The fully connected structure is fed with all the parameters and outputs the vector containing the first ten natural frequencies $f_i$ of the top plate.

The train and test sets are found randomly shuffling the datasets' elements and dividing them with a ratio of 9/1.
The network is then trained using the Levenberg-Marquardt training function, which is a combination of gradient descent and Newton's method and a variable number of epochs for each training, always below 100.   
All the networks we train have a coefficient of multiple determination $R^2$ that is greater than 0.9, so that we can consider our method as reliable and use it in the following studies \cite{gonzalez2021data}.

\section{Optimization procedure}

Once the neural network is trained, we have a multidimensional function that depends on the outline shape, thickness profile and material parameters of the plate. We call this function 
\begin{equation}
    \mathscr{F}(p_1,...,p_{20};t_1,...,t_8;m_1,...,m_7) = f_i~.
\end{equation}
$\mathscr{F}$ is continuous and differentiable in all its variables, but not necessarily invertible.  
We can use this function in two different ways. In the first, we start from the parameters of a given violin mesh and compute its eigenfrequency values. This application is useful to predict the vibrational response of a top plate before it is built. In the second method, we set a vibrational feature $f_0$ on the natural frequencies vector. Then, we optimize the input parameters of $\mathscr{F}$ to find a mesh whose response $f_i$ is as close as possible to $f_0$. In this paper, we focus on the second method, optimizing the shape of violin plates in several cases.
To measure how well the estimated eigenfrequency set $f_i$ fits with the desired vibrational feature $f_0$, we define several error functions $\epsilon$ that can vary according to the analyses and we minimize it.
We do so using the Matlab \textit{fminsearch} function, which follows the Nelder-Mead simplex method \cite{lagarias1998convergence}.
We consider a bounded minimization version of the algorithm to avoid large variations of the parameters that could lead to unrealistic violin plates. We set a maximum change of the variables of 20\% in both directions, consistent with how we built the dataset.
During the optimization process, the error function is evaluated for a maximum of $200*N_v$ times before the minimum is reached, where $N_v$ is the number of variables considered.

\section{Optimizing the outline of the violin top plate}

In this section we focus on the outline shape of the plate analyzing its role in determining several vibrational features.
The vibrational characteristics that we consider are taken as examples to show the capability of our method in improving the violin design while we do not discuss their importance in the instrument-making field \cite{davis2013effective}.
Any other feature could be chosen and the geometry of the violin plate could be optimized for that.

\begin{figure}
\centering
\includegraphics[height=65mm]{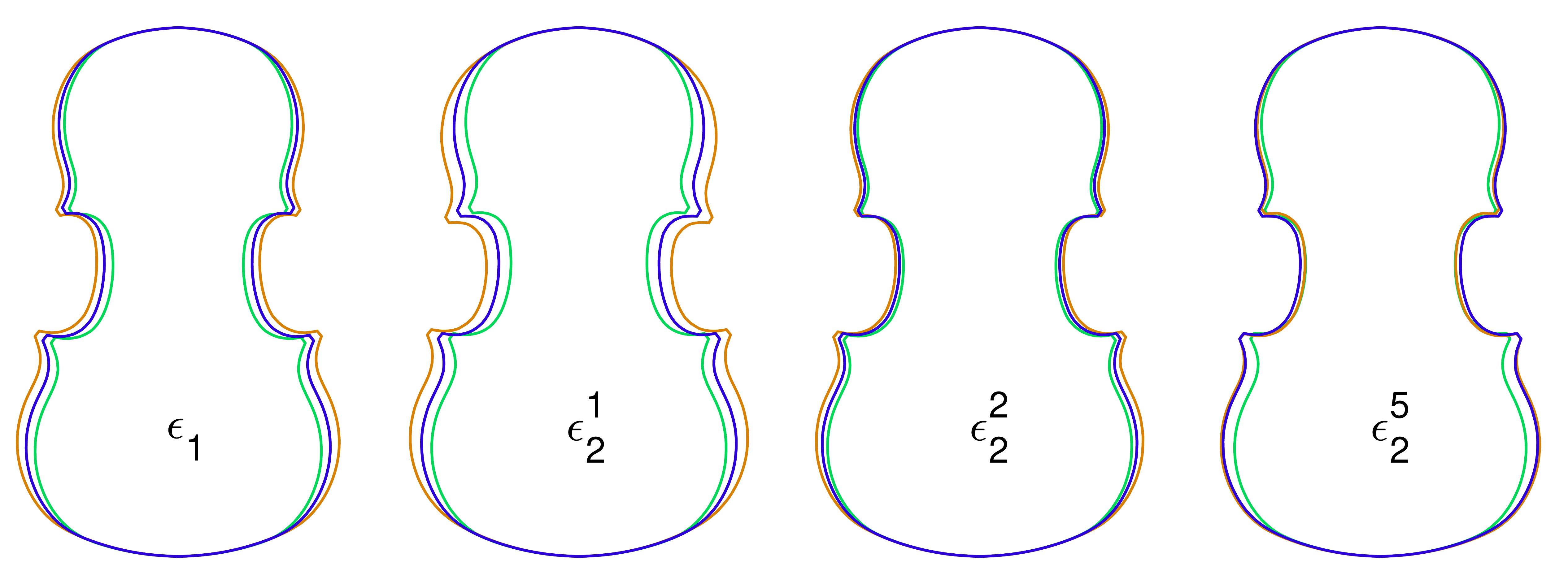}
\caption{Outline shapes optimized for different values of $f_{52}$ and single modes $f_i$. The $f_{52}$ ratio is optimized by $\epsilon_1$, while the single modes $i$ are optimized by $\epsilon_2^i$.
Orange and green lines represent the outlines for increased and decreased ratio/eigenfrequency values, respectively.}
\label{NN_optimizedOutline}
\end{figure}

The vibrational feature we consider in this first study is the ratio between the fifth and second modes of the plate, namely $f_{52} = f_5/f_2$, which is a simple modal relation that violin makers usually compute in the violin building process.
The role of this ratio has been discussed in \cite{curtin2005tap}.
We start by defining the loss function as
\begin{equation}
   \epsilon_1 = (\alpha-f_{52})^2
\label{epsilon1}
\end{equation}
where $\alpha$ is the value of the frequency ratio we want to achieve, and we minimize the loss function for changing values of $\alpha$. We start from $\alpha=2.3$, which is the optimal value set in the literature, and we vary it by 5\% in both directions.
We look for the minimum of the loss function by changing the outline parameters and predicting the frequency values using the neural network.
Our starting point are the parameters that best fit the reference violin we used to build the dataset, which has a $f_{52}^{\textrm{ref}}= 2.57$.
The resulting outlines are shown in Fig.~\ref{NN_optimizedOutline} with the $\epsilon_1$ label.
The blue line represents the reference outline, while 
the orange and green ones are obtained respectively by increasing and decreasing $\alpha$ from the nominal value.
There is a correlation between the value of the ratio and the outline shape and, in particular, the ratio increases as we increase the width of the violin.
To validate the results we take the mesh of the plate whose predicted value of the ratio is equal to $f_{52}^{\textrm{pred}} = 2.30$ and we simulate it in Comsol with FEM to actually compute the ratio, since its set of parameters was not part of the dataset. The resulting value of the ratio is $ f_{52}^\textrm{{FEM}} = 2.31$. We have checked all the outlines, and the error has always been smaller than $1\%$.

Secondly, we consider the first vibrational modes of the violin top plate and we change its outline shape to optimize the frequency value of each of them.
In particular, we modify the eigenfrequencies by 5\% in both directions from their reference values.
This study aims to see how changes in the outline shape affect the frequency values of individual vibrational modes and detect whether some modes are more subject to change than others.

The error function we consider in this case is
\begin{equation}
    \epsilon_2^i = (\beta-f_{i})^2
    \label{epsilon2}
\end{equation}
where $f_i$ is the eigenfrequency corresponding to the $i$-th mode and $\beta$ the target value. In particular $\beta = f_i^{\mathrm{ref}} \pm 5\%$.

The results of this study are shown in Fig.~\ref{NN_optimizedOutline}, where we have plotted only the outlines of modes 1, 2 and 5, which are the most observed by luthiers in the construction processes of their violins.
In the Figure, the blue line represents the reference outline, while 
the orange and green ones are obtained respectively by increasing and decreasing the respective eigenfrequency values.
We observe that not all the modes behave in the same way. Still, the relationship between each individual vibrational mode and the outline shape is unique and different from the others, meaning that the relationship between the violin geometry and its vibrational properties is highly non-linear. In particular, mode 1 shows a large outline change, contrary to mode 2, where the profile remains almost unchanged. 

\section{Vibratory equivalence between changes in outline shape and thickness distribution}

Here we want to see if, from the vibrational point of view, a change in the outline shape of a violin top plate can be balanced with a variation in its thickness distribution and vice versa. The purpose of this study is to prove that these two geometric aspects, under certain circumstances, can be equivalent to each other in determining the final vibratory response of the plate.

We start from the geometry and the eigenfrequency values of the reference violin. Then, we impose a change in the outline shape of the plate and optimize its thickness distribution to bring the eigenfrequency values back to their initial values. We also do the opposite by imposing a change in the thickness distribution and optimizing the outline shape to balance the shift.
The whole procedure can be summarized as:
\begin{gather*}
    f_i \rightarrow \mathscr{F}(p + \delta, t) \rightarrow \mathscr{F}(p + \delta , t') \rightarrow f_0 ~,\\
    f_i \rightarrow \mathscr{F}(p, t + \delta) \rightarrow \mathscr{F}(p' , t + \delta) \rightarrow f_0 ~,
\end{gather*}
where $f_0$ has to be as close as possible to $f_i$.
We modify the outline and thickness parameters randomly, as $p_i' \rightarrow p_i(1+\delta_i)$ where $\delta_i$ is taken from a zero-mean Gaussian distribution with increasing $\sigma$ values.
At the end of each optimization, we measure the error between the obtained eigenfrequencies and the reference ones. We do so by using two different error functions, which are
\begin{equation}
    \epsilon_3 = \frac{1}{N} \sum_{i=1}^{N}\left( \frac{\left|f^{opt}_i-f^{ref}_i\right|}{f^{ref}_i} \right) \qquad ~,~
    \epsilon_4    = \frac{\left|  \bar{f}_i^{opt} -\bar{f}_i^{ref} \right|}{\bar{f}_i^{ref}}
    \label{epsilon4}
    \end{equation}

where $f^{opt}_i$ and $f^{ref}_i$ are vectors containing the predicted values of the first 10 eigenfrequencies of the optimized and reference plates respectively and $\bar{f}_i$ is the mean of a frequency vector.
The first error function considers the distances between all the individual modes in the optimized and target cases. On the other hand, the second function measures only the one between the mean frequencies in the two cases.
$\epsilon_3$ represents the ideal error function and aims to make the response of the optimized plate identical to the target one.
Nevertheless, we are aware that this function is difficult to minimize due to the strongly non-linear relationships present between the vibrational modes. For this reason, we have introduced $\epsilon_4$, which is less severe and less difficult to minimize but at the same time produces meaningful results.

\begin{figure}
\centering
\includegraphics[height=65mm]{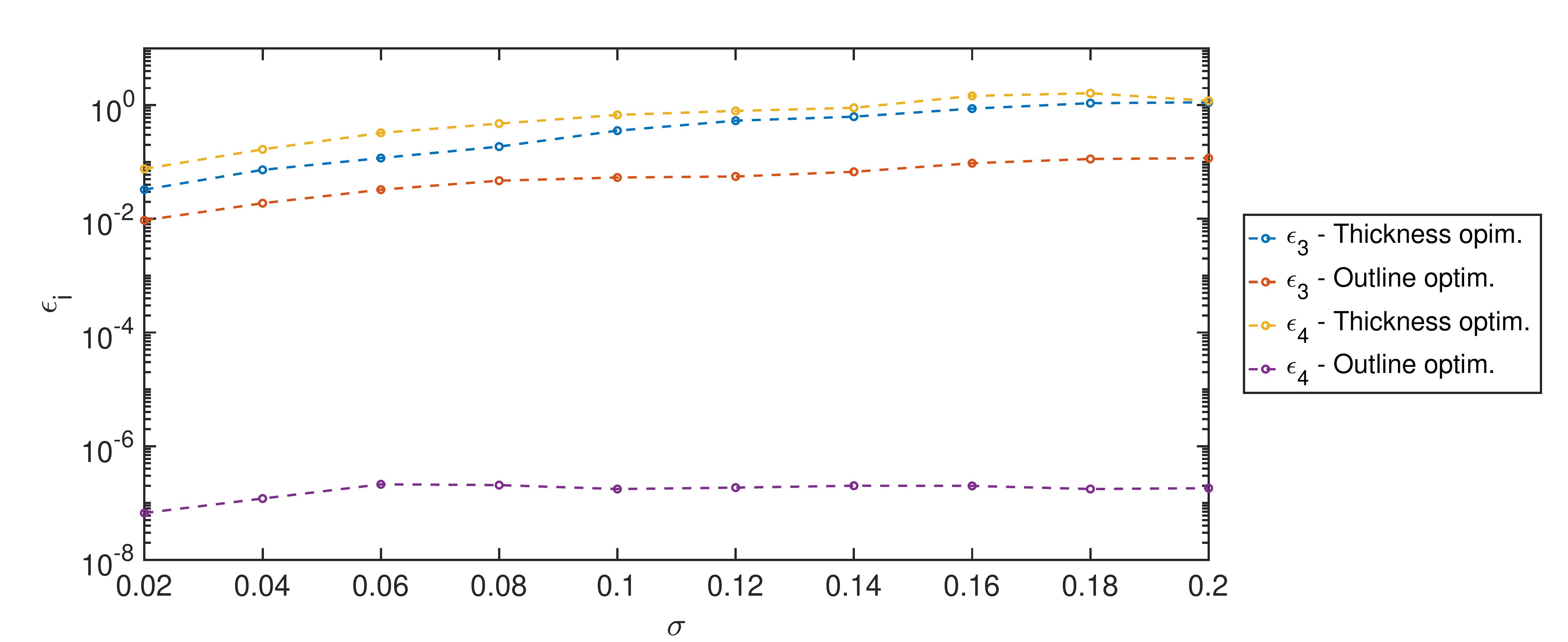}
\caption{Outline (thickness) optimization for changing values of the thickness (outline) parameters. Here are shown the values of $\epsilon_3$ and $\epsilon_4$ for increasing values of $\sigma$. The plotted values are the average of 20 different simulations.}
\label{outline_vs_thickness}
\end{figure}

Figure \ref{outline_vs_thickness} shows the results of the analysis for increasing values of $\sigma$, obtained as an average between the values of 20 different optimizations.
We observe that the best results are given by the outline optimizations. This means that we can compensate for variations in the thickness distribution by changing the outline shape, but not vice versa.
This could be due to the greater number of parameters needed to define the contour leading to a more efficient higher-dimensional optimization.
However, the thickness optimization manages to achieve acceptable results, especially for low $\sigma$, with $\epsilon_i$ values that do not exceed 10\%.
Furthermore, as we anticipated, we have better results for $\epsilon_4$ than for $\epsilon_3$. This is evident in the outline case, while for the thickness the two error functions have similar behavior.

Our initial hypothesis of vibrational equivalence between variations in the outline and thickness distribution has proved wrong. We have shown that it is possible to optimize the vibrational response of a plate through the outline shape, while the thickness distribution can be used for fine-tuning.

\section{Shape optimization for material properties changes}

In this section we add to the analysis also the wood properties of the plate. In particular, we impose a change in the material parameters and modify the geometry of the plate to keep its vibrational response unaltered.
This is done because the wood selection is a critical step in building a musical instrument. We want to show that it is possible to work on the geometry to optimize the final vibrational response notwithstanding the material properties, which is potentially groundbreaking for violin makers.

The material properties we can control in $\mathscr{F}$ are density, Young's modulus, and Poisson's ratio in the 3 main directions of wood, for a total of 7 parameters. In this study, we randomly modify all of them,  as $m_i' \rightarrow m_i(1+\delta_i)$ where $\delta_i$ is taken from a zero-mean Gaussian distribution with $\sigma = 0.2$, values somehow representative of the actual variation in tone woods. Then, we optimize the geometry of the plate in 3 different ways to bring its eigenfrequency values back to the reference ones considering $\epsilon_3$ as error function. The three optimizations are thickness only $\mathscr{F}(p_i;\bar{t}_i;m_i')$, outline only $\mathscr{F}(\bar{p}_i;t_i;m_i')$, and thickness and outline together $\mathscr{F}(\bar{p}_i;\bar{t}_i;m_i')$ where the bar indicates which parameters we are optimizing.

\begin{figure}
\centering
\includegraphics[height=65mm]{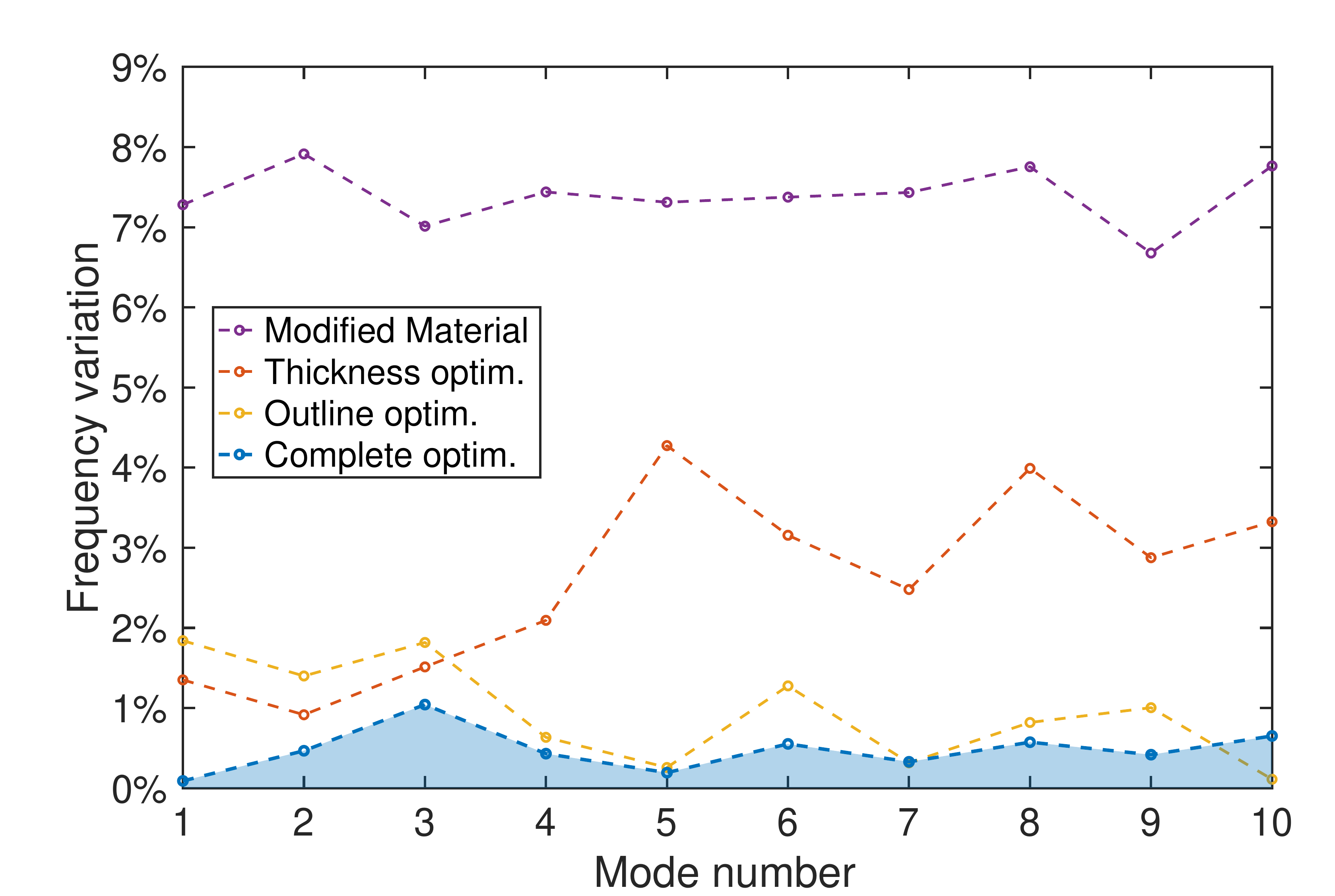}
\includegraphics[height=65mm]{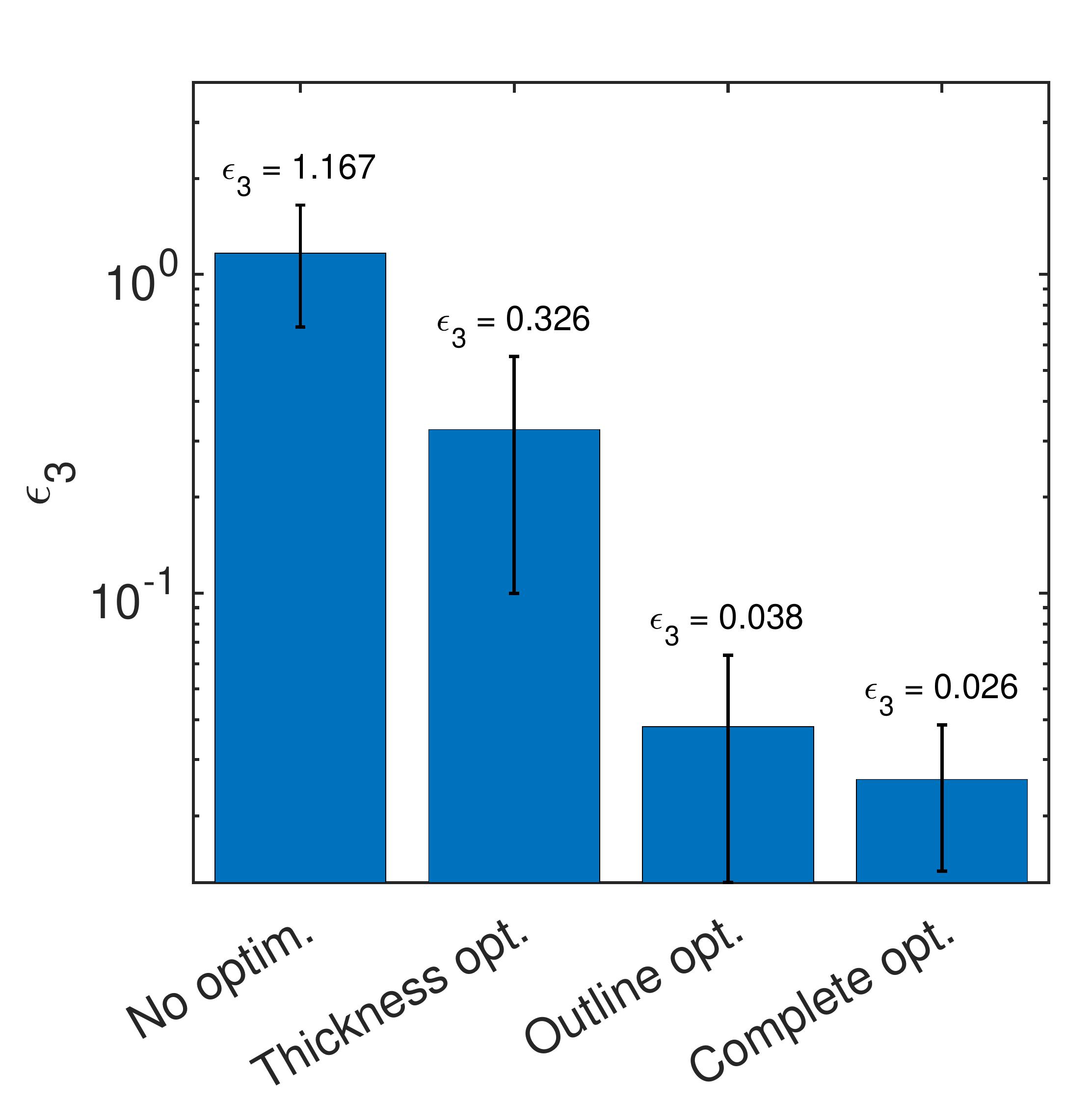}
\caption{Left: Eigenfrequency values obtained modifying the material properties of the plate and optimizing its shape for $\epsilon_3$. All the frequencies are normalized with respect to the reference ones. Right: Error $\epsilon_3$ in the four cases considered, logarithmic scale. The error bars are the stardard deviation from the mean value.}
\label{optimization_complete}
\end{figure}

To highlight the variations, the left plot in Fig.~\ref{optimization_complete} shows the optimized eigenfrequency values normalized by the reference ones. The plotted values are the average of 20 different optimizations.
We obtain the best overall results by optimizing both the outline shape and the thickness distribution. The full optimization gives for all $f_i$ the best results, save in the case of mode 10. Notably, the $f_1$ can only be predicted with less than 1\% error for the complete optimization. Notice how the optimization lowers the error almost two orders off magnitude.

Among all the mechanical parameters, we now focus on two of the most relevant: density ($\rho$) and Young's modulus in the wood's principal direction ($E_y$). If we approximate the wood plate to a bar we obtain that their ratio has clear physical meaning
\begin{equation}
    c = \sqrt{\frac{E_y}{\rho}} ~,
    \label{wave_speed}
\end{equation}
where $c$ is longitudinal wave propagation velocity in the plate \cite{norton1990fundamentals} (disregarding the effect of the Poisson's ratio).
We changed the density and Young's modulus values in the range [-10\%, +10\%] with a step of 2\%, optimizing the violin shape both in thickness and outline to obtain the minimum values of $\epsilon_3$.
The starting values are the standard parameters of Sitka Spruce, that are $\rho = 400 kg/m^3$ and $E_y = 10.8 GPa$.

\begin{figure}
\centering
\includegraphics[height=65mm]{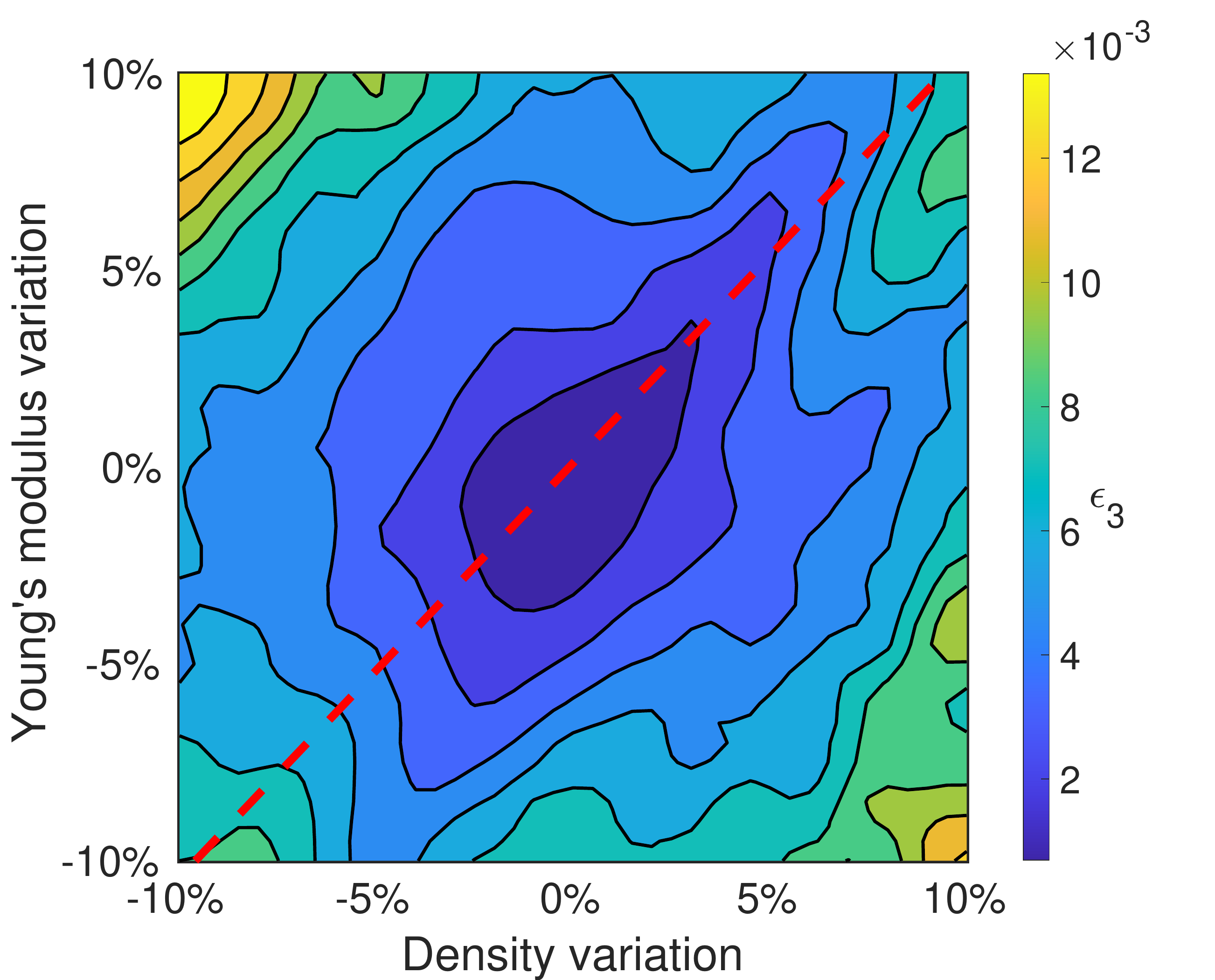}
\includegraphics[height=65mm]{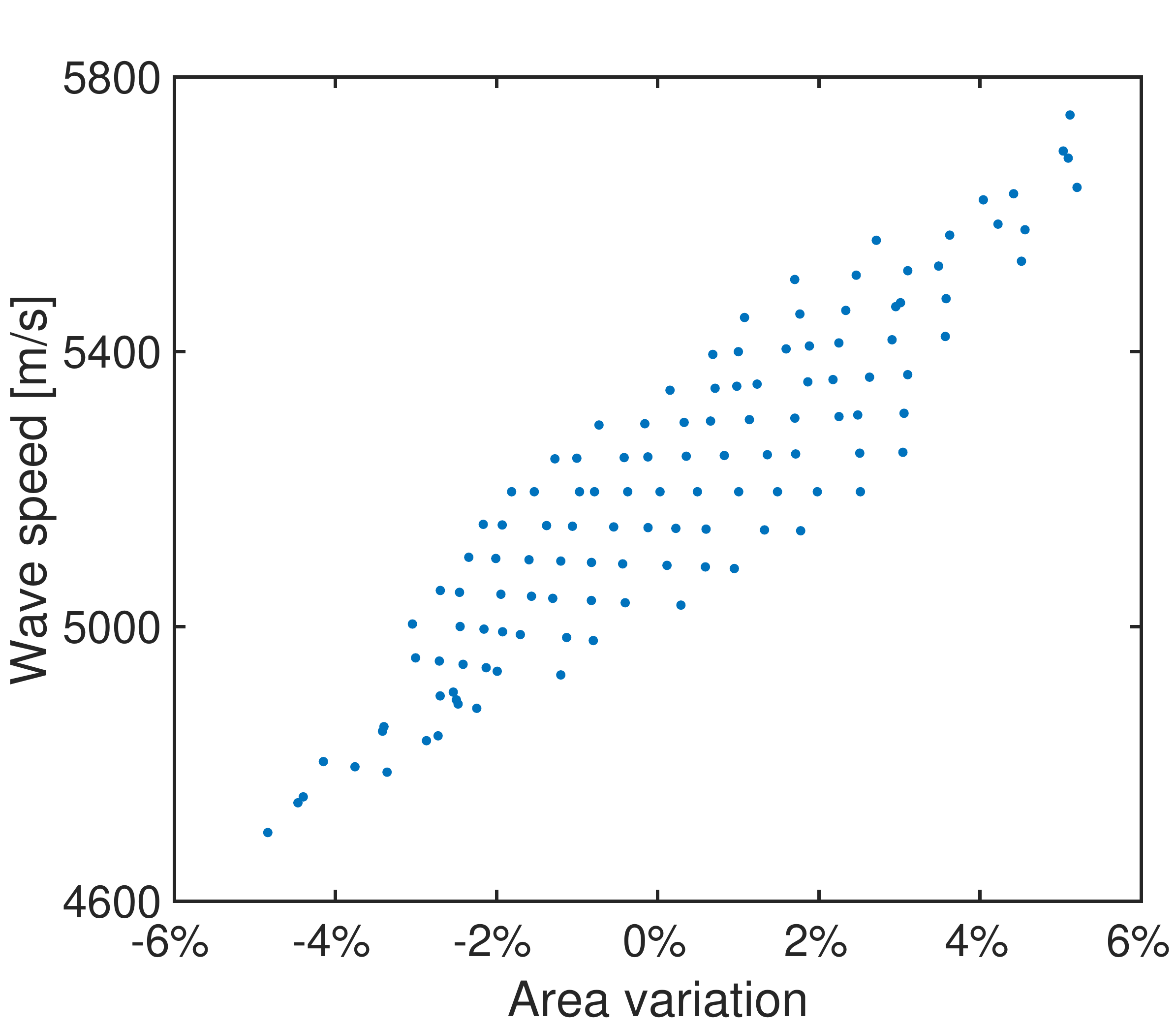}
\caption{Left: Geometry optimization of $\epsilon_3$ as density and Young's modulus in the wood principal direction change. The red line represents a constant sound speed region, where the optimization achieves better results. Right: Scatter plot of the area variation before/after optimization (normalized by the area of the reference violin) versus $c$ for the points in the grid of the left figure, $R^2 = 0.834$.}
\label{material_opt}
\end{figure}

The image on the left of Fig.~\ref{material_opt} shows the results of the analysis, where the value of $\epsilon_3$ increases as we increase the variation of the material parameters. The red line represents a region where the sound speed is constant and equal to $c = 5200 m/s$, and the optimization achieves better results. Interestingly, our algorithm is able to find a rather good optimal shape for rather large variations of the density and stiffness. It also seems to indicate that the larger the sound speed difference between two samples, the `harder' (larger $\epsilon_3$) is to optimize the shape. To quantify the shape variation, in the right plot of Fig.~\ref{material_opt}, we show the area change versus the sound speed variation for each point of the left image. The surface of the plate and the wave speed are highly correlated ($R^2 = 0.834$), which intuitively tells us that the lower the sound speed of a material, the narrower the violin needs to be to vibrate as the reference model. On the contrary, if the sound speed is higher, the violin shape needs to wider. The variation is rather significant and one can recognize distinct historical examples on the resulting outlines.  

\section{Conclusions}
We have presented a new approach to geometric optimization of violin top plates with FEM simulations and Neural Networks. Rather than a priori deciding on what to optimize, we have shown that doing a Gaussian sampling of the parameter space can be used to predict an arbitrary loss function based on the eigenfrequency values. The computational time needed to create the dataset is comparable to the time needed for one optimization, whereas the speed up gained with the NN is almost 3 orders of magnitude.

The results presented here point towards a complete re-thinking of today's violin-making practice: rather than copy old models violin to reproduce their sound, violin makers should look at what type of material we have nowadays and modify the shape of the violin to obtain a desired vibrational response. We have shown that varying only the thickness of the violin, as traditional plate tuning suggests \cite{hutchins1981acoustics}, is far from optimal, and varying the outline is much more effective. In future developments of this study we aim to increase the number of parameters used to define the violin plate to make the model as complete as possible. We can consider different instrument lengths or different profiles for the longitudinal and transverse archings. Varying these parameters will change the vibrational response of the plates, but we do not see any reason why the same methodology cannot be applied for those cases.

We have used our method to compute the vibrational response of violin plates, but the same methodcould probably work in either other geometries or other physical variables (e.g.~ displacement at a point, stiffness of the resulting geometry). We have not studied the prediction of the spatial behavior of the modes, but research in our own group has shown that this can be predicted with convolutional and autoencoders networks \cite{olivieri2020nahcnn}. The path towards optimization of sound seems finally within reach, more than 300 years after Stradivari built the instruments that inspired this study.

\bibliographystyle{icsv_bib}
\bibliography{references}

\end{document}